\documentclass{jetpl}
\usepackage[utf8]{inputenc}
\usepackage[english,english]{babel}
\twocolumn
\usepackage{textcomp}
\lat

\title{A new type of charge-density-wave pinning in 
orthorhombic TaS$_3$ crystals with quenching defects}

\rtitle{}

\sodtitle{}

\author{V.\,E.\,Minakova, A.\,M.\.Nikitina, 
S.\,V.\,Zaitsev-Zotov\/\thanks{e-mail: SerZZ@cplire.ru}}

\rauthor{V.\,E.\,Minakova, A.\,M.\.Nikitina, S.\,V.\,Zaitsev-Zotov}

\sodauthor{Minakova, Nikitina, Zaitsev-Zotov}

\address{Kotel'nikov Institute of Radioengineering and Electronics of 
Russian Acamemy of Sciences, Mokhovaya 11, bld.7, 125009 Moscow, Russia}

\dates{\today}{*}

\abstract{Diminishing in the 
concentration of quenching defects during thermocycling of orthorhombic TaS$_3$ samples 
in the temperature range below the Peierls transition temperature $T <T_P$ is observed. 
It makes it possible to study the character of pinning of the 
charge density wave (CDW) by these defects.
A number of fundamental differences from pinning by ordinary local pinning centers — 
impurities and point defects — have been found.
We conclude that quenching defects are extended (non-local) 
objects (presumably, dislocations) that can diffuse from the crystal during 
low-temperature thermоcycling due to their strong interaction with the CDW, 
which is intrinsic for Peierls conductors.
The presence of these defects leads to a previously unknown non-local type of 
CDW pinning that acts on $T_P$ and the threshold field for the onset of the CDW sliding, $E_T$, differently in comparison with the local pinning centers.}


\begin{document}

\maketitle

{\bf Introduction.} 
It is known that the interaction of the CDW with impurities and defects leads to 
its pinning \cite{cdwreview1,cdwreview2, Larkin1, Larkin2, Lee1}. 
As a result, the current-voltage characteristics of Peierls conductors remain 
linear until the electric field exceeds the threshold value $E_T$ 
corresponding to the onset of the CDW sliding.
$E_T$ depends on the concentration of pinning centers, $n$. 
The growth of $n$ not only increases $E_T$ and smears the transition to the
nonlinear conductance but also leads to a decrease of
$T_P$ and broadening of the Peierls transition.
This behavior has been well studied for local pinning centers, such as impurities 
and point defects. The theory usually considers two simplest cases of pinning 
by local centers — strong ($E_T \propto n$) and weak 
($E_T \propto n^2$) \cite{Fukuyama, Lee2}
\footnote{Real pinning can contain both weak and strong pinning elements \cite{Abe1,Abe2}.}.
Strong pinning centers are introduced by irradiating the sample with neutrons 
or fast electrons, leading to the displacement of lattice atoms 
\cite{Mihaly1, Mihaly2, Mutka, Petristchev}, or by doping the crystal with 
charged impurities such as Ti. In this case, the change in the temperature 
of the Peierls transition $\Delta T_P \propto \sqrt{n}$. 
Weak pinning centers appear under non-ideal conditions of synthesis 
\cite{Petristchev} or with the introduction of uncharged impurities, for example, 
isoelectronic \cite{Pei-Ling}, in this case $\Delta T_P \propto n$. 
As a result, $\Delta T_P \propto \sqrt{E_T}$ as in the case of 
weak and strong pinning.

Defects in the crystal can be created in another way - by sharp cooling (quenching) 
of the crystals during synthesis.
As a rule, these are dislocations of different types \cite {Wilke, Kittel}.
To get rid of them, high-temperature annealing is usually used. 
In the Peierls conductors, quenching defects were not previously intentionally 
created and not studied.

In this work, we have studied the CDW pinning by quenching defects in 
orthorhombic TaS$_3$ ({\it o}-TaS$_3$) crystals. 
As it turned out, the presence of a strong interaction of the CDW with these defects 
causes a change in their concentration during low-temperature thermocycling 
and leads to the evolution of the transport properties of the crystals under study.
Due to this evolution, which is manifested in the modification of the Ohmic and 
nonlinear conductance with an increase in the number of thermocyclings, $N$, 
we were able to study these dependences and compare the results with those 
known for the pinning types studied earlier.
The comparison showed that the character of the CDW pinning by quenching defects 
is different than in the case of local pinning carried out by 
impurities and point defects.

{\bf Experimental.} 
{\it o}-TaS$_3$ is a Peierls conductor with a single transition at $T_P\approx 220$~K 
and an energy gap crossing entire Fermi surface.
When $E<E_T$ its Ohmic conductance at $T_P/2  \lesssim T < T$ follows the activation 
law with the activation energy 850~K \cite{cdwreview1}.
In crystals with a low content of impurities and defects $E_T \lesssim 1$~V/cm.

{\it o}-TaS$_3$ crystals were grown by the gas transport technology in a temperature gradient 
of 650-670~\textcelsius\ for 3 days
in an evacuated (up to $10 ^ {- 5}$ Torr) and sealed quartz ampoule 
from a mixture of Та (99,995~\%) and S (99.999~\%) 
in stoichiometric proportions together with the transport agent (10~\% excess S). 
Defects were created at the end of the synthesis by rapid quenching from growth 
to room temperature.

To eliminate the influence of size effects on $T_P$ and $E_T$  
\cite{zzreview}, relatively large crystals with transverse sizes of  
2 -- 10~$\mu$m and lengths of 1.8 -- 3~mm  were studied.
Contacts with a resistance of $\sim 10$~Ohms and a width of $\lesssim 80$~$\mu$m 
were made by cold soldering of indium. 

The values of $T_P$ and $E_T$ -- the main crystal-quality characteristics -- 
were extracted from the temperature dependence of the Ohmic conductance, $G(T)$, 
measured in the voltage-controlled regime at $V_{dc } \ll V_T$, 
and the electric field dependence of conductance, $G\equiv I/V$, respectively. 
All measurements were carried out in a two-contact configuration. 
The dependences of $G(T)$ were measured with cooling at a rate of 2~K/min 
in the range from 300~K to 77~K, in some cases down to $T = 8$~K. 
In all measurements, the accuracy of stabilization and temperature measurement 
was better than 0.1~K.
$T_P$ corresponds to the maximum of the temperature 
dependence of the logarithmic derivative of the resistance with respect to the 
inverse temperature, $d\ln R/d(1/T)$.
$E_T$ was determined as the onset of a weak nonlinearity of the dependence $G(E)$ 
at $T~=~90$~K after applying a voltage $V> V_{T}$ 
to remove the metastability of the CDW \cite{cdwreview1}.
For two shortened samples, $E_T$ was also measured at $T = 77$~K and $T = 130$~K.

{\bf Results and discussions.} 
Testing of samples from seven different growth batches subjected to quenching 
showed that two types of crystals coexist in the same batch:
\begin{enumerate}
\item Crystals with stable properties - no changes in $T_P$ and $E_T$ 
during repeat measurements, with $T_P \approx$ 208 -- 213~K.
\item Unstable crystals ($\sim 10$~\%) -- with initially extremely low $T_P$, 
usually in the range of 195~--~200~K, and a significant increase in $T_P$ 
and a decrease in $E_T$ with each new measurement.
\end{enumerate}
This unusual behavior observed in the second case indicates an improvement in 
the crystals quality due to the weakening of the CDW pinning caused by  
thermocycling.
It was suggested that the different degree of stability of the properties 
of the grown crystals is due to the different content of quenching defects in them, 
and the effect of weakening the CDW pinning is due to their ability 
to move and exit the sample.
\begin{figure}
\resizebox{80mm}{!}{\includegraphics{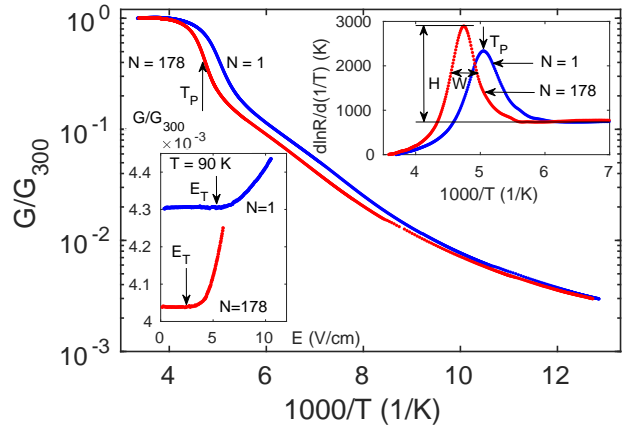}}
\caption{Fig. 1. (Color online) Temperature dependences of the Ohmic conductance 
$G$, normalized to the conductance at room temperature, $G_{300}$, at the first and 178th 
thermocycling for the sample No. 1. 
The top inset shows methods for determining both the value of $T_P$ from the maximum 
of the temperature dependence of $d\ln R/d(1/T)$ and the values of the height, $H$, 
and of the width, $W$, of the peak $d\ln R/d(1/T)$. 
The bottom inset shows electric field dependences of the conductance, $G(E)/G_{300}$,
for the same sample at $N=1$ and $N=178$.}
\label{Fig.1}
\end{figure}
This assumption is based on the analogy with the results of \cite{Gill, Venera}, 
devoted to the study of the kinetics of In impurities and its influence 
on the transport properties of the Peierls conductors 
\footnote{Note that the kinetics of defects in the Peierls conductors 
has not been studied previously.}.
In both papers, In atoms intercalated into the crystal from contacts 
upon heating in an inert gas flow, and then diffused into the sample.
In Ref. \cite{Gill} in NbSe$_3$ a time-dependent change in $E_T$ 
was observed after the onset of the CDW sliding, which was explained 
by the redistribution of In impurities in the sample, 
caused by their interaction with the CDW.
An activation dependence of the In diffusion coefficient in NbSe$_3$ was found. 
In \cite{Venera} In diffusion into the contact region of the {\it o}-TaS$_3$ sample 
led to changes in the photoconductance spectra, 
a decrease in the value of $T_P$ ($\Delta T_P \approx 5$~K) 
and an increase in $E_T$ about 20~\%.

Fig.~\ref{Fig.1} shows the dependence of $G(T)$, measured during the first 
cooling of sample No. 1 to $T~=~77$~K (blue curve), together with the dependence 
measured at $N = 178$ (red curve). 
The top inset shows the temperature dependence of the derivative 
of the resistance  $d\ln R/d(1/T)$, 
demonstrating that as a result of the 178 thermocyclings, the Peierls transition 
became more abrupt, and $T_P$ increased noticeably. 
The bottom inset shows the difference between the dependences of $G(E)$ at $N = 1$ 
and $N = 178$. 
It is seen that with an increase in $N$ the transition to a nonlinear state 
becomes sharper, and $E_T$ decreases.

\begin{figure}
\includegraphics[width=8.8 cm]{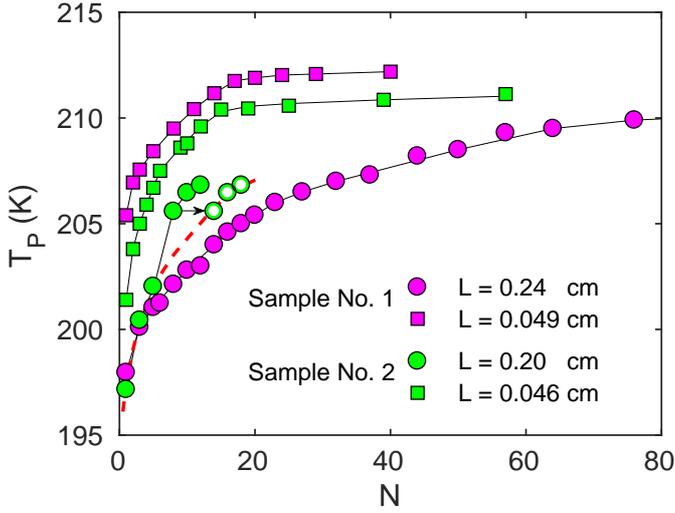}
\caption{Fig. 2. (Color Online) Change in the shape of the $T_P(N)$ dependence 
when the sample is shortened. The red dashed line with circles 
is a dependence that would have been with standard measurements (details in the text).}
\label{Fig.2}
\end{figure}

\begin{figure}
\includegraphics[width=8.8 cm]{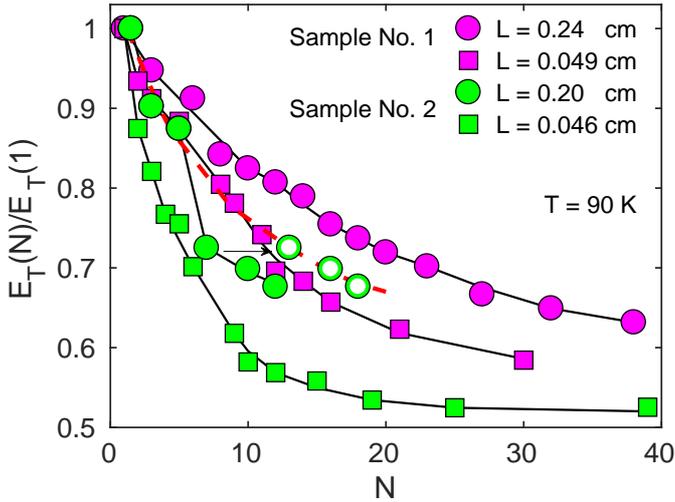}
\caption{Fig. 3. (Color online) Change in the shape of $E_T(N)$ dependences 
normalized to $E_T(1)$, with decreasing $L$. 
The red dashed line with circles is a dependence that would have been with 
standard measurements (details in the text).}
\label{Fig.3}
\end{figure}

The dependences of $T_P(N)$ and $E_T(N)$ for sample No. 1 are shown in 
Fig.~\ref{Fig.2} and~\ref{Fig.3}, respectively (pink circles).
The greatest changes in both quantities occur when $N$ is small.
Since with the growth of $N$  these changes became smaller,
some thermocyclings were carried out without measurements,
note that, an increase in $T_P$ was observed even in the absence of an electric field 
$E \ll E_T$ on the sample.
As $N$ grows, the curves tend to saturate, and $T_P$
reaches values characteristic of initially stable samples.

The total growth of $T_P$  due to thermocyclings for some samples reaches 15~K, 
and $E_T$ decreases by 2.5~times. 
From the initial and final value of $E_T$, we can roughly estimate the initial, $n_0$, 
and final, $n$, effective concentration of all defects and impurities in the sample 
(as if they all were ordinary defects providing weak pinning), 
as well as a change in the concentration, $\Delta n$, 
due to the exit of unstable defects: $n = 2.3~\cdot~10^{-4}$~at.\%,     
$\Delta n = 1.1 \cdot 10^{-4}$~at.\%. 
It means that almost a third of all impurities and defects initially present 
in our samples disappeared due to thermocyclings.
The result is surprising -- it was previously reported that low-temperature 
thermocycling increases the degree of defectiveness of the sample 
\cite{cdwreview1, Nad'}.

If the observed weakening of the CDW pinning is caused by the exit of 
quenching defects during thermocycling, then an increase in $N$ should lead 
to their non-uniform distribution in the sample volume, with a lower concentration 
at the edges of the crystal.
Since the process of exit of defects from anisotropic materials occurs mainly 
in the direction of the main axis \cite{Wilke}, 
then we can expect its acceleration with decreasing sample length.

To test this hypothesis, two additional contacts were made in the central part 
of sample No. 1.
For $N=178$ the value of $T_P$ for the entire length of the sample, $L = 0.24$~cm, 
was $211.0$~K, and in the central part with $L_{new} = 0.049$~cm, $T_P=205.3$~K. 
Consequently, there are really more defects in the middle of the sample 
than at the ends.
After that, the sample was cut just near the central contacts, and further measurements 
were performed on the shortened sample.

Fig.~\ref{Fig.2} allows us to compare the $T_P(N)$ dependences of the long and 
short versions of sample No. 1, as well as of both versions of sample No. 2, 
which was subjected to the same shortening procedure.
Points with $N = 1$ for short samples were measured before they were cut.
For sample No. 2, $T_{Pnew}$ is also 5.4~K lower than the final $T_P$ for long sample.
It can be seen that $T_P$ in the short sample No. 1 grows faster than in the long one. 
For the sample No. 2, this fact is not so obvious. This is due 
to the fact that the rate of $T_P$ and $E_T$ changes also depends on the lowest 
temperature of thermocycling and increases with its decrease. For the 
long sample~No.~2, before the fourth measurement of $T_P$ ($N = 8$), there were 
two intermediate thermocyclings ($N = 6$ and $7$) not down to $T = 77$~K, 
but down to $T = 8$~K. They turned out to be more effective than standard ones --
their effect is equivalent to eight thermocyclings down to $T = 77$~K.
Red dashed line with circles on it is drawn with this fact in mind.

Fig.~\ref{Fig.3} shows normalized dependences of $E_T(N)$ 
for long and short versions of both samples.
The values of $E_T(1)$ for short samples, as expected, turned out to be larger 
than the corresponding final values of $E_T$ for long samples.
For the long sample~No. 2, a sharp kink in the dependence is visible
between  the 3rd and 4th measurements, caused by intermediate thermocyclings 
down to $T = 8$~K. 
Accounting for their greater efficiency gives a smooth relationship (red dashed line), 
which allows us to see that $E_T$, as well as $T_P$, changes faster in short samples.

A detailed study of the process of the exit of quenching defects caused by thermocycling 
shows:
\begin{enumerate}
\item Changes occur mainly during thermocycling in the $T \lesssim T_P$ region.
\item The effect of thermocycling at $T> T_P $ is small or completely absent.
\item Long storage at $T < T_P = const$ even in the presence of an electric field 
$E > E_T$ (processes occurring without changing the CDW wave vector  
\cite{Wang, Inagaki} and the coefficient of elasticity of the CDW) 
do not have a significant effect on the defect exit process.
\end{enumerate}

Thus, we can conclude that the process of the exit of quenching defects 
is most intense when the CDW state changes -- during thermocycling 
in the Peierls transition region where the CDW is formed, and especially 
in the low $T$ region, 
where the configuration of the CDW changes greatly due to changes 
in the wave vector value and its elastic properties caused by changes in screening conditions.
Therefore, it can be assumed that the cause of the elimination of defects 
from a crystal is the presence of the CDW and its strong interaction 
with pinning centers \cite{Gill}.

\begin{figure}
\resizebox{80mm}{!}{\includegraphics{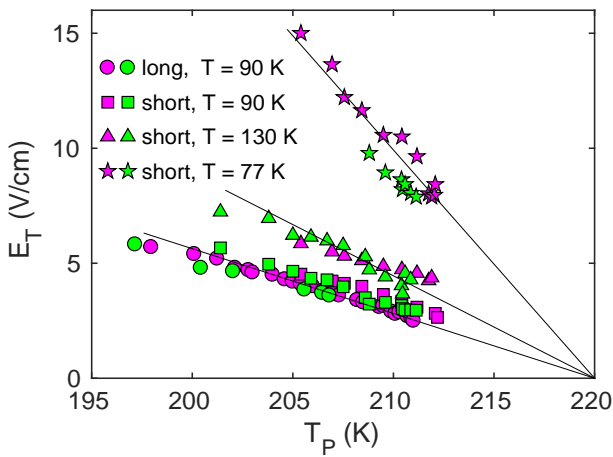}}
\caption{Fig. 4. (Color online) Dependence $E_T(T_P)$ for the long and short versions
of samples No. 1 (pink icons) and 2 (green icons) for different $T$.}
\label{Fig.4}
\end{figure}

Fig.~\ref{Fig.4} shows the relationship between the values of $T_P$ and $E_T$ 
with the same $N$ for long and short versions of both samples.
Let us first consider the case of long samples (pink and green circles).
Although the rates of change of $T_P$ and $E_T$ on $N$ for the samples 
are very different, for seven of the eight studied
long samples the dependences $E_T(T_P)$ are almost the same 
(for the eighth sample, the dependence shifted slightly upwards).
This dependence is close to linear and crosses the temperature axis at $T = 220$~K,
coinciding with $T_P$ for pure {\it o}-TaS$_3$ crystals.

The dependences $E_T(T_P)$ for short samples are also measured at $T = 77$ 
and 130~K.
At $T = 90$~K, they are slightly ($\approx $ by 1~V/cm) shifted upwards 
compared to long samples.
The shift is caused by the somewhat overestimated value of $E_T$, arising 
due to the two-probe measurement configuration \cite {Zybtsev},
which is significant with decreasing $L$ and increasing  $T$. 
The dependences of $E_T(T_P)$ of short samples 
are linear and, taking this shift into account, also converge at $T=220$~K. 
Their slope depends non-monotonously on $T$ and has a minimum at $T=90$~K 
(this is a consequence of the dependence  $E_T(T)$). 

We also tried to describe the results obtained by the law 
$\sqrt{E_T} \propto \Delta T_P$, inherent in local pinning (both strong \cite {Mutka}, 
and weak \cite {Pei-Ling}). 
This method gave a new point at which all lines converge at $E_T \rightarrow 0$: 
$T \approx 234$~K, which is much higher than $T_P$. 
This result is not physical.
Similar examination of the law $E_T^2 \propto \Delta T_P$ 
gave the convergence point of all lines $T = 215$~K. 
This case, in principle, may occur taking into account the fact that the presence 
of other impurities and defects can actually reduce the transition temperature to 
$T~\simeq~215$~K, which we see for the best stable samples without quenching defects.
In any case, we can conclude: pinning by quenching defects is described by a law 
different from the $\sqrt{E_T} \propto \Delta T_P$ law, 
which is characteristic of local pinning \cite{Mutka,Pei-Ling}. 

The linear relationship $\Delta T\propto E_T$ can be attempted to explain 
taking into account the fact that pinning at high temperatures 
near the Peierls transition may be weak 
($\Delta T_P\propto n_i$), and at low temperatures -- strong ($E_T\propto~n_i$). 
Arguments in favor of changing the type of pinning with decreasing 
$T$ were discussed in a number of papers \cite{Tucker,ZZRM}, 
and the minimum of $E_T(T)$ dependence at $T_{min}~\approx~90~\div 100$~K was 
associated with this change.
However, in our case such an explanation is impossible, since the dependence 
$\Delta T\propto E_T$ occurs already at $T=130$~K, which is higher than 
$T_{min}$ ($E_{T_{130}}> E_{T_{90}}$).

It remains to assume that the observed difference in the relationship between $E_T$ 
and $T_P$ in the studied samples from the previously known 
$\sqrt{E_T} \propto \Delta T_P$ law is due to a fundamentally different, 
new type of CDW pinning, which was not previously studied theoretically, 
neither experimentally.
It is probably caused by extended (non-local) objects (for example, dislocations).
The assumption is also based on the fact that it is dislocations that most often 
appear during quenching and are able to move relatively easily across 
the crystal \cite{Wilke, Kittel}. 
But the main argument in favor of this assumption is the observed feature 
of the new pinning described below: 
the non-local pinning does not have a significant effect on the Peierls state.

As we see from the upper inset in Fig.~\ref{Fig.1}, 
the magnitude of the maximum 
$d\ln R/d(1/T)$ and the width of the peak for our samples change slightly 
despite of a significant change in $T_P$.
The opposite result was observed for all types of local pinning.
For instance, in studies of the effect of fast electron irradiation on {\it o}-TaS$_3$, 
an increase in the defects concentration 
(displacements of Ta atoms and vacancies left by them, which are centers 
of strong pinning) led to a significant suppression \cite{Petristchev} 
or to an almost complete disappearance \cite{Mutka} of the maximum $d\ln R/d(1/T)$ 
with the same change in $T_P$.
A similar picture was observed on samples with weak pinning caused by 
growth point defects arising under nonideal synthesis conditions \cite{Petristchev}  
or isoelectronic impurities in {\it o}-TaS$_3$ alloys with Nb \cite{Pei-Ling}.
In the case of a non-uniform distribution of weak pinning centers created by 
In diffusion into the contact area \cite{Venera}, 
the rate of suppression of the maximum $d\ln R/d(1/T)$ 
was also greater than in our (also not completely uniform) case.

\begin{figure}
\includegraphics[width=8.0 cm]{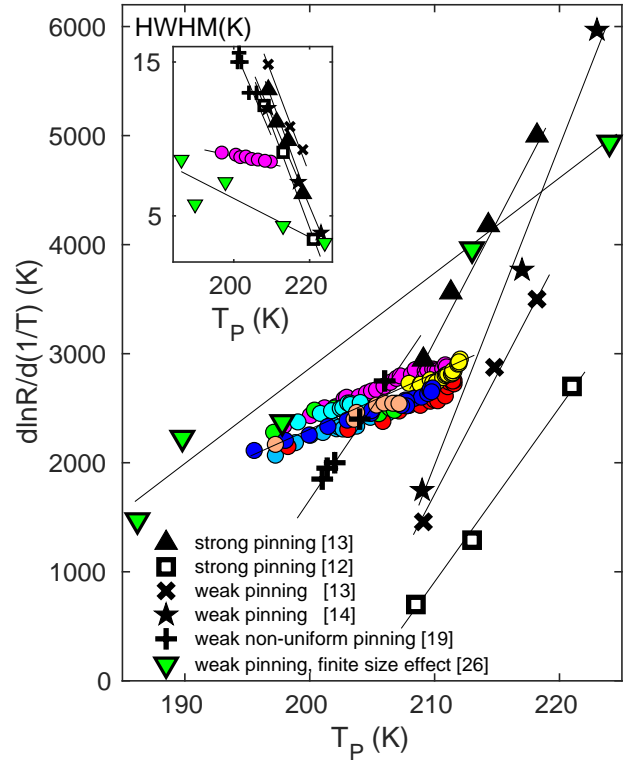}
\caption{Fig. 5. (Color online) Relationship between the maximum value of $d\ln R/d(1/T)$ 
and the value of $T_P$ for all samples with quenching defects (colored circles), 
for samples with different types of local pinning (black icons) 
and samples with small sections (green triangles).
The inset shows the relationship between the values of $HWHM$ and $T_P$ 
for the same samples.}
\label{Fig.5}
\end{figure}

The relationship between the maximum value of $d\ln R/d(1/T)$ and $T_P$ 
for all our samples is shown in Fig.~\ref{Fig.5} (colored circles) 
along with similar dependences for samples with local pinning, 
obtained on the basis of the data from the above-mentioned researches (black icons), 
as well as dependences for samples with small sections $S < 0.1$~$\mu$m$^2$
(green triangles, data from \cite{Borodin}). 
In all the cases of local pinning, the maximum is suppressed much faster than in our case.
With a decrease in $S$ in the case of weak local pinning, the rate of change of 
$d\ln R/d(1/T)$ with $T_P$ decreases, but it still remains higher than 
in our study.
Consequently, CDW disordering at the Peierls transition is significant in the 
presence of local pinning and much weaker in non-local pinning.

Another characteristic of the degree of sharpness of the transition is 
the half-width of the peak $d\ln R/d(1/T)$, measured at its half-maximum,  
$HWHM$. 
The inset to Fig.~\ref{Fig.5} shows the relationship of the values of $HWHM$ 
and $T_P$ for sample No.~1 along with all the above-mentioned cases of local pinning 
and samples with small $S$. 
And again, the change in $HWHM$ in the case of the non-local pinning is much 
weaker than in the local pinning cases.
In \cite{Mutka} a considerable smearing of 
the Peierls transition due to irradiation of the sample was accompanied by 
the broadening of the sharp low-temperature satellites and their 
transformation into continuous diffuse planes 
in electron diffraction measurements.
The results were associated with the loss of the CDW of transverse coherence, 
which is typical for the high-temperature region.  
Apparently, non-local defects break the crystal in the transverse direction 
into separate domains, along which the CDW almost does not lose longitudinal 
coherence and remains three-dimensional ordered.
Thus, the interaction of the CDW with non-local defects is fundamentally different than in the case of local pinning.

In conclusion, we summarize the main results. 
Unusual features of the CDW pinning by quenching defects in {\it o}-TaS$_3$ samples 
were found, namely:
\begin{enumerate}
\item Pinning is unstable and is eliminated during thermocycling 
in the temperature range $T < T_P$.
\item With an increase in the number of low-temperature thermocyclings, 
pinning becomes spatially nonuniform with a lower concentration of 
defects at the ends of the crystal.
\item Pinning is described by a law different from the $\sqrt{E_T} \propto \Delta T_P$ 
law, which is characteristic of local pinning.
\item The discovered pinning is less destructive for the Peierls state than 
the local one.
\end{enumerate}

The presence of these features let us suggest that quenching defects 
are macroscopic (non-local) objects.
They lead to a previously unknown type of CDW pinning with properties different 
from local pinning ones.
Such the CDW pinning can be caused, for example, by dislocations.
And the strong interaction of CDW with any defects (specificity of the 
Peierls conductor), as well as the ability of dislocations to move through the crystal, 
lead to their forced diffusion and exit from the crystal during low-temperature 
thermocycling.

\end{document}